\documentclass[12pt]{article}
\usepackage{lscape}
\usepackage{citesort}
\usepackage{amssymb}
\usepackage{appendix}
\usepackage{multirow}

\begin{document}
\def\qq{\langle \bar q q \rangle}
\def\uu{\langle \bar u u \rangle}
\def\dd{\langle \bar d d \rangle}
\def\sp{\langle \bar s s \rangle}
\def\GG{\langle g_s^2 G^2 \rangle}
\def\Tr{\mbox{Tr}}
\def\figt#1#2#3{
        \begin{figure}
        $\left. \right.$
        \vspace*{-2cm}
        \begin{center}
        \includegraphics[width=10cm]{#1}
        \end{center}
        \vspace*{-0.2cm}
        \caption{#3}
        \label{#2}
        \end{figure}
	}
	
\def\figb#1#2#3{
        \begin{figure}
        $\left. \right.$
        \vspace*{-1cm}
        \begin{center}
        \includegraphics[width=10cm]{#1}
        \end{center}
        \vspace*{-0.2cm}
        \caption{#3}
        \label{#2}
        \end{figure}
                }

\def\ds{\displaystyle}
\def\beq{\begin{equation}}
\def\eeq{\end{equation}}
\def\bea{\begin{eqnarray}}
\def\eea{\end{eqnarray}}
\def\beeq{\begin{eqnarray}}
\def\eeeq{\end{eqnarray}}
\def\ve{\vert}
\def\vel{\left|}
\def\ver{\right|}
\def\nnb{\nonumber}
\def\ga{\left(}
\def\dr{\right)}
\def\aga{\left\{}
\def\adr{\right\}}
\def\lla{\left<}
\def\rra{\right>}
\def\rar{\rightarrow}
\def\lrar{\leftrightarrow}  
\def\nnb{\nonumber}
\def\la{\langle}
\def\ra{\rangle}
\def\ba{\begin{array}}
\def\ea{\end{array}}
\def\tr{\mbox{Tr}}
\def\ssp{{\Sigma^{*+}}}
\def\sso{{\Sigma^{*0}}}
\def\ssm{{\Sigma^{*-}}}
\def\xis0{{\Xi^{*0}}}
\def\xism{{\Xi^{*-}}}
\def\qs{\la \bar s s \ra}
\def\qu{\la \bar u u \ra}
\def\qd{\la \bar d d \ra}
\def\qq{\la \bar q q \ra}
\def\gGgG{\la g^2 G^2 \ra}
\def\GG{\langle g_s^2 G^2 \rangle}
\def\g5{\gamma_5 \not\!q}
\def\x{\gamma_5 \not\!x}
\def\g5{\gamma_5}
\def\sb{S_Q^{cf}}
\def\sd{S_d^{be}}
\def\su{S_u^{ad}}
\def\sbp{{S}_Q^{'cf}}
\def\sdp{{S}_d^{'be}}
\def\sup{{S}_u^{'ad}}
\def\ssp{{S}_s^{'??}}

\def\sig{\sigma_{\mu \nu} \gamma_5 p^\mu q^\nu}
\def\fo{f_0(\frac{s_0}{M^2})}
\def\ffi{f_1(\frac{s_0}{M^2})}
\def\fii{f_2(\frac{s_0}{M^2})}
\def\O{{\cal O}}
\def\sl{{\Sigma^0 \Lambda}}
\def\es{\!\!\! &=& \!\!\!}
\def\ap{\!\!\! &\approx& \!\!\!}
\def\ar{&+& \!\!\!}
\def\arrr{\!\!\!\! &+& \!\!\!}
\def\ek{&-& \!\!\!}
\def\vev{&\vert& \!\!\!}
\def\kek{\!\!\!\!&-& \!\!\!}
\def\cp{&\times& \!\!\!}
\def\se{\!\!\! &\simeq& \!\!\!}
\def\eqv{&\equiv& \!\!\!}
\def\kpm{&\pm& \!\!\!}
\def\kmp{&\mp& \!\!\!}
\def\mcdot{\!\cdot\!}
\def\erar{&\rightarrow&}

% .........................................................

\def\simlt{\stackrel{<}{{}_\sim}}
\def\simgt{\stackrel{>}{{}_\sim}}

% .........................................................

\title{
         {\Large
                 {\bf
Magnetic moments of $J^P = {3\over 2}^-$ baryons in QCD
                 }
         }
      }

\author{\vspace{1cm}\\
{\small T. M. Aliev \thanks {
taliev@metu.edu.tr}~\footnote{Permanent address: Institute of
Physics, Baku, Azerbaijan.}\,\,,
M. Savc{\i} \thanks
{savci@metu.edu.tr}} \\
{\small Physics Department, Middle East Technical University,
06531 Ankara, Turkey }}

\date{}

\begin{titlepage}
\maketitle
\thispagestyle{empty}

\begin{abstract}

The magnetic moments of the low lying, negative parity, spin-3/2 baryons are
calculated within the light cone QCD sum rules method. The contributions
coming from the positive parity, spin-3/2 baryons, as well as from the
positive and negative parity spin-1/2 baryons are eliminated by
constructing combinations of various invariant amplitudes corresponding to
the coefficients of the different Lorentz structures.

\end{abstract}

%\vspace{1cm}
~~~PACS numbers: 11.55.Hx, 13.40.Gp, 14.20.Gk

\end{titlepage}

\section{Introduction}

The study of the structure and properties of hadrons represents one of the
main directions in particle and nuclear physics. There are different
approaches in investigation of the structure of hadrons, especially, the
most promising one in this direction is the study of the radiative decays
and electromagnetic properties of the hadrons and their excitations (for a
review, see for example \cite{Rkul01}). Negative parity partners of the
octet baryons arise from increasing of orbital angular momentum by one unit.

The investigation of the magnetic moment of the negative parity baryons can
provide us useful information about their internal structure. Note that the
study of the magnetic moment of the $N^\ast$ baryon has already been planned
at Mainz Microtron (MAMI) facility \cite{Rkul02,Rkul03,Rkul04} and 
Jefferson Laboratory (JLAB) \cite{Rkul05}. The main
difficulty in measurement of the magnetic moments of the excited baryons is
can be attributed to the considerably large width they posses. Due to this
fact the magnetic moments can be measured from the polarization observables
of the decay products of the excited resonances. The very first effort in
measuring the magnetic moment of the $\Delta^{+}(1232)$ in the reaction
$\gamma p \to p \pi^0 \gamma$ has been realized at MAMI, and future program
to get information about the magnetic moment of the $N^\ast(1535)$
baryon in the reaction $\gamma p \to p \eta \gamma$ has already been planned.     
The magnetic moments of the spin-1/2 negative parity baryons have been
calculated within the nonrelativistic quark model \cite{Rkul06}, simple quark
model  \cite{Rkul07}, lattice QCD  \cite{Rkul08}, chiral perturbation theory
\cite{Rkul09}, light cone QCD sum rules (LCSR) \cite{Rkul10}, and effective
Hamiltonian model \cite{Rkul11}.

In the present work we calculate the magnetic moments of the negative
parity, spin-3/2 partners of the octet baryons in framework of the LCSR
method (more about LCSR formalism can be found in \cite{Rkul12}). It should
be reminded here that, theoretically there exists only one work where
magnetic moment of the negative parity, spin-3/2 $\Lambda$-baryon is calculated
in framework of the chiral quark model \cite{Rkul13}.

The paper is organized as follows. In section 2 we introduce the
interpolating currents for the spin-3/2 partners of the octet baryons and
derive sum rules for the magnetic moments of negative parity baryons. In the
same section, we also calculate the two-point correlators in order to
obtain the sum rules for mass and residues of the negative parity, spin-3/2
Section 3 is devoted to numerical analysis of the obtained sum rules for the
magnetic moments of these baryons, and discussion and conclusion.

\section{Calculation of the magnetic moments of negative parity, spin-3/2
baryons from LCSR}

The starting point in calculation of the magnetic moment of the
negative parity, spin-3/2 baryons from the QCD side is the consideration of the
following correlation function:
\bea
\label{ekul01}
\Pi_{\mu\nu} = i \int d^4x e^{ipx} \lla 0 \vel \mbox{\rm T} \Big\{ \eta_\mu^B
(x) \bar{\eta}_\nu^B(0) \Big\}\ver 0 \rra_F~,
\eea
where $\eta_\mu^B$ is the interpolating current for the corresponding
baryon, $F$ refers to the external electromagnetic field, and T is the time
ordering operator.

In order to obtain the sum rules for the magnetic moments, the correlation
function is calculated in terms of hadrons and the quark-gluon degrees of
freedom. Using then the quark-hadron duality ansatz these two forms of the
correlation function are related by using the analytical continuation.

The phenomenological part of the correlation function can be obtained by
saturating it with the complete states of single particle hadronic states
carrying the same quantum numbers as the interpolating currents. 
We get from Eq.(\ref{ekul01}):
\bea
\label{ekul02}
\Pi_{\mu\nu} \es {\la 0 \ve \eta_\mu \ve \mbox{\small${3\over 2}^+$}(p_2)
\ra \la \mbox{\small${3\over 2}^+$}(p_2) \gamma (q) \ve
\mbox{\small${3\over 2}^+$}(p_1)  \ra \la \mbox{\small${3\over
2}^+$}(p_1) \ve \bar{\eta}_\nu \ve 0 \ra  
\over (m_{\mbox{\tiny${3\over 2}^+$}}^2 - p_1^2)
(m_{\mbox{\tiny${3\over 2}^+$}}^2 - p_2^2)} \nnb \\
\ar {\la 0 \ve \eta_\mu \ve \mbox{\small${3\over 2}^-$}(p_2)
\ra \la \mbox{\small${3\over 2}^-$}(p_2) \gamma (q) \ve
\mbox{\small${3\over 2}^+$}(p_1)  \ra \la \mbox{\small${3\over
2}^+$}(p_1) \ve \bar{\eta}_\nu \ve 0 \ra
\over (m_{\mbox{\tiny${3\over 2}^+$}}^2 - p_1^2)
(m_{\mbox{\tiny${3\over 2}^-$}}^2 - p_2^2)} \nnb \\
\ar {\la 0 \ve \eta_\mu \ve \mbox{\small${3\over 2}^+$}(p_2)
\ra \la \mbox{\small${3\over 2}^+$}(p_2) \gamma (q) \ve
\mbox{\small${3\over 2}^-$}(p_1)  \ra \la \mbox{\small${3\over
2}^-$}(p_1) \ve \bar{\eta}_\nu \ve 0 \ra
\over (m_{\mbox{\tiny${3\over 2}^-$}}^2 - p_1^2)
(m_{\mbox{\tiny${3\over 2}^+$}}^2 - p_2^2)} \nnb \\
\ar{\la 0 \ve \eta_\mu \ve \mbox{\small${3\over 2}^-$}(p_2)
\ra \la \mbox{\small${3\over 2}^-$}(p_2) \gamma (q) \ve
\mbox{\small${3\over 2}^-$}(p_1)  \ra \la \mbox{\small${3\over
2}^-$}(p_1) \ve \bar{\eta}_\nu \ve 0 \ra
\over (m_{\mbox{\tiny${3\over 2}^-$}}^2 - p_1^2)
(m_{\mbox{\tiny${3\over 2}^-$}}^2 - p_2^2)}~.
\eea

In general the current $\eta_\mu$ interacts not only with spin-3/2 states,
but also with spin-1/2 states. Since our aim is to calculate the magnetic
moments of the negative parity, spin-3/2 baryons, the contributions coming
spin-1/2 baryons must be eliminated. The prescription to eliminate these
unwanted contributions can be summarized as follows.

The general form of the matrix element of $\eta_\mu$ between spin-1/2 and
vacuum states can be written as,
\bea
\label{ekul03}
\la 0 \vel \eta_\mu \ver \mbox{\small${1\over 2}$}(p) \ra = (A p_\mu + B
\gamma_\mu) u(p)~.
\eea
Multiplying both sides with $\gamma^\mu$, and using condition $\eta_\mu
\gamma^\mu=0$ we get, 
\bea
\label{ekul04}
\la 0 \vel \eta_\mu \ver \mbox{\small${1\over 2}^+$} (p) \ra =
B \Bigg( - {4 \over m_{\mbox{\tiny${1\over 2}^+$}} } p_\mu +
\gamma_\mu \Bigg) u(p)~,
\eea
and similarly,
\bea 
\label{ekul05}
\la 0 \vel \eta_\mu \ver \mbox{\small${1\over 2}^-$} (p) \ra =
B \gamma_5 \Bigg( - {4 \over m_{\mbox{\tiny${1\over 2}^-$}} } p_\mu +
\gamma_\mu \Bigg) u(p)~.
\eea
We see from Eqs. (\ref{ekul04}) and (\ref{ekul05}) that the unwanted
spin-1/2 state contributions are proportional to either $p_\mu$ or
$\gamma_\mu$, and they must be eliminated. For this goal an ordering
procedure of the Dirac matrices is needed which is necessary
in obtaining the independent structures. In the present study the ordering
of the Dirac matrices chosen is $\gamma_\mu \rlap/{\varepsilon}
\rlap/{q} \rlap/{p} \gamma_\nu$.
In the light of these remarks we can now proceed to calculate $\Pi_{\mu\nu}$
given in Eq. (\ref{ekul02}) in terms of hadrons, for which the following
matrix elements are needed,
\bea
\label{ekul06}
\la 0 \vel \eta_\mu \ver \mbox{\small${3\over 2}^+$} (p) \ra \es 
\lambda_{\mbox{\tiny${3\over 2}^+$}} u_\mu (p)~, \nnb \\
\la 0 \vel \eta_\mu \ver \mbox{\small${3\over 2}^-$} (p) \ra \es 
\lambda_{\mbox{\tiny${3\over 2}^-$}} \gamma_5 u_\mu (p)~, \nnb \\
\la \mbox{\small${3\over 2}^+$}(p) \vel \bar{\eta}_\nu \ver 0 \ra \es
\lambda_{\mbox{\tiny${3\over 2}^+$}} \bar{u}_\nu (p)~, \nnb \\
\la \mbox{\small${3\over 2}^-$}(p) \vel \bar{\eta}_\nu \ver 0 \ra \es
- \lambda_{\mbox{\tiny${3\over 2}^-$}} \bar{u}_\nu (p) \gamma_5~,
\eea
where $u_\mu(p)$ is the Rarita-Schwinger spinor for the spin-3/2 particles.
The matrix element $\la B_{\mbox{\tiny${3\over 2}^+$}} (p_2) \ve
B_{\mbox{\tiny${3\over 2}^+$}} (p_1) \ra_\gamma$ is determined in the
following way:
\bea
\label{ekul07}
\la B_{\mbox{\tiny${3\over 2}^+$}} (p_2) \ve            
B_{\mbox{\tiny${3\over 2}^+$}} (p_1) \ra_\gamma \es
\varepsilon_\rho \bar{u}_\alpha (p_2) {\cal O}^{\alpha\rho\beta} u_\beta
(p_1) \equiv \nnb \\
&&\varepsilon_\rho \bar{u}_\alpha (p_2) \Bigg\{-g^{\alpha\beta} \Bigg[
\gamma^\rho (f_1^+ +f_2^+) +
{(p_1+p_2)^\rho \over 2 m_{\mbox{\tiny${3\over 2}^+$}} } f_2^+ + q^\rho f_3^+
\Bigg] \nnb \\
\ek {q^\alpha q^\beta \over 2 m_{\mbox{\tiny${3\over 2}^+$}}^2 }
\Bigg[ \gamma^\rho (g_1^+ +g_2^+) + {(p_1+p_2)^\rho \over 2
m_{\mbox{\tiny${3\over 2}^+$}} } g_2^+ + q^\rho g_3^+ \Bigg] \Bigg\} u_\beta
(p_1)~,
\eea
where $f_i$ and $g_i$ are the form factors whose values are needed only at
the point $q^2=0$. The matrix elements $\la B_{\mbox{\tiny${3\over 2}^-$}}
(p_2) \ve B_{\mbox{\tiny${3\over 2}^-$}} (p_1) \ra_\gamma$ and
$\la B_{\mbox{\tiny${3\over 2}^+$}} (p_2) \ve B_{\mbox{\tiny${3\over 2}^-$}}
(p_1) \ra_\gamma$
can be obtained
from Eq. (\ref{ekul07}) by making the replacements of the form factors as,
\bea
\label{nolabel01}
f_i \to f_i^\ast, ~~~g_i \to g_i^\ast,~\mbox{\rm and}~
f_i \to f_i^{tr}, ~~~g_i \to g_i^{tr}, \nnb
\eea
respectively. For the $3^+/2 \rar 3^-/2$ transition the matrix element such
as given in Eq. (\ref{ekul07}) should contain $\gamma_5$ matrix before
the Rarita-Schwinger spinor $u_\beta (p)$, which follows from the parity
consideration. 
Note that for the real photons the terms multiplying $f_3$
and $g_3$ can be neglected since $\varepsilon_\rho q^\rho = 0$. Summation
over spin-3/2 states is performed according to the following formula:
\bea
\label{ekul08}
\sum_s u_\mu (p,s) \bar{u}_\alpha (p,s) = - (\rlap/{p} + m) \Bigg[
g_{\mu\alpha} - {1\over 3} \gamma_\mu \gamma_\alpha - {2 p_\mu p_\alpha
\over 3 m^2} + {p_\mu \gamma_\alpha - p_\alpha \gamma_\mu \over 3 m}
\Bigg]~.
\eea  
It should be remembered again that the terms proportional to
$\gamma_\mu$ on the left and those to $\gamma_\alpha$ on the right; as well as
to $p_{1\alpha}$ and $p_{2\mu}$ contain contributions from spin-1/2 states
(see Eqs. (\ref{ekul04}) and (\ref{ekul05})), and therefore can be neglected.
After elimination of the spin-1/2 states, the only structure that
contains contributions from spin-3/2 states is $g_{\mu\alpha}$. As a result, the
hadronic part of the correlation function containing only spin-3/2
contributions can be written as,
\bea
\label{ekul09}
\Pi_{\mu\nu} \es
{\lambda_{\mbox{\tiny${3\over 2}^+$}}^2 \over (m_{\mbox{\tiny${3\over
2}^+$}}^2-p_1^2) (m_{\mbox{\tiny${3\over 2}^+$}}^2-p_2^2)} 
(\rlap/{p}_2 + m_{\mbox{\tiny${3\over
2}^+$}}) g_{\mu\alpha} \varepsilon_\rho \Bigg\{-g^{\alpha\beta} \Bigg[
\gamma^\rho (f_1^+ +f_2^+) +
{(p_1+p_2)^\rho \over 2 m_{\mbox{\tiny${3\over 2}^+$}} } f_2^+ \Bigg] \nnb \\
\ek {q^\alpha q^\beta \over 2 m_{\mbox{\tiny${3\over 2}^+$}}^2}
\Bigg[ \gamma^\rho (g_1^+ +g_2^+) + {(p_1+p_2)^\rho \over 2
m_{\mbox{\tiny${3\over 2}^+$}} } g_2^+ \Bigg] \Bigg\} (\rlap/{p}_1 +
m_{\mbox{\tiny${3\over 2}^+$}}) g_{\nu\beta} \nnb \\
\ek {\lambda_{\mbox{\tiny${3\over 2}^+$}} \lambda_{\mbox{\tiny${3\over
2}^-$}}\over (m_{\mbox{\tiny${3\over 2}^-$}}^2-p_1^2)
(m_{\mbox{\tiny${3\over 2}^+$}}^2-p_2^2)}
(\rlap/{p}_2 + m_{\mbox{\tiny${3\over 2}^+$}}) g_{\mu\alpha}
\varepsilon_\rho \Bigg\{-g^{\alpha\beta} \Bigg[
\gamma^\rho (f_1^{tr} +f_2^{tr}) +
{(p_1+p_2)^\rho \over m_{\mbox{\tiny${3\over 2}^+$}} +
m_{\mbox{\tiny${3\over 2}^-$}} } f_2^{tr} \Bigg] \nnb \\
\ek {q^\alpha q^\beta \over 2 m_{\mbox{\tiny${3\over 2}^+$}}^2}
\Bigg[ \gamma^\rho (g_1^{tr} +g_2^{tr}) + {(p_1+p_2)^\rho \over 
m_{\mbox{\tiny${3\over 2}^+$}} + m_{\mbox{\tiny${3\over 2}^-$}} } g_2^{tr} \Bigg]
 \Bigg\} \gamma_5 (\rlap/{p}_1 + m_{\mbox{\tiny${3\over 2}^-$}})
\gamma_5 g_{\nu\beta} \nnb \\
\ar  {\lambda_{\mbox{\tiny${3\over 2}^-$}} \lambda_{\mbox{\tiny${3\over
2}^+$}}\over (m_{\mbox{\tiny${3\over 2}^+$}}^2-p_1^2)
(m_{\mbox{\tiny${3\over 2}^-$}}^2-p_2^2)}
\gamma_5 (\rlap/{p}_2 + m_{\mbox{\tiny${3\over 2}^-$}}) g_{\mu\alpha}
\varepsilon_\rho \Bigg\{-g^{\alpha\beta} \Bigg[
\gamma^\rho (f_1^{{tr}^\ast} +f_2^{{tr}^\ast}) +
{(p_1+p_2)^\rho \over m_{\mbox{\tiny${3\over 2}^+$}} +
m_{\mbox{\tiny${3\over 2}^-$}} } f_2^{{tr}^\ast} \Bigg] \nnb \\
\ek {q^\alpha q^\beta \over m_{\mbox{\tiny${3\over 2}^+$}}^2}
\Bigg[ \gamma^\rho (g_1^{{tr}^\ast} +g_2^{{tr}^\ast}) + {(p_1+p_2)^\rho \over 
m_{\mbox{\tiny${3\over 2}^+$}} + m_{\mbox{\tiny${3\over 2}^-$}} } g_2^{{tr}^\ast} \Bigg]
 \Bigg\} \gamma_5 (\rlap/{p}_1 + m_{\mbox{\tiny${3\over 2}^+$}})
g_{\nu\beta} \nnb \\
\ek {\lambda_{\mbox{\tiny${3\over 2}^-$}}^2 \over (m_{\mbox{\tiny${3\over
2}^-$}}^2-p_1^2) (m_{\mbox{\tiny${3\over 2}^-$}}^2-p_2^2)} 
\gamma_5 (\rlap/{p}_2 + m_{\mbox{\tiny${3\over
2}^-$}}) g_{\mu\alpha} \varepsilon_\rho \Bigg\{-g^{\alpha\beta} \Bigg[
\gamma^\rho (f_1^- +f_2^-) +
{(p_1+p_2)^\rho \over 2 m_{\mbox{\tiny${3\over 2}^-$}} } f_2^- \Bigg] \nnb \\
\ek {q^\alpha q^\beta \over 2 m_{\mbox{\tiny${3\over 2}^-$}}^2}
\Bigg[ \gamma^\rho (g_1^- +g_2^-) + {(p_1+p_2)^\rho \over 2
m_{\mbox{\tiny${3\over 2}^-$}} } g_2^+ \Bigg] \Bigg\} (\rlap/{p}_1 +
m_{\mbox{\tiny${3\over 2}^-$}}) \gamma_5 g_{\nu\beta}~.
\eea

It is shown in \cite{Rkul14} that, in the nonrelativistic limit, in the
presence of external uniform magnetic field the maximum
energy of the baryon is equal to $3 (f_1+f_2) B$, where $B$ is the magnitude
of the field. In other words, $3 (f_1+f_2)$ is equal to the magnetic moment
at $q^2=0$. Therefore, among many structures we chose the one which
multiply the coefficient $(f_1+f_2)$. Using this fact, the hadronic part of
the correlation function given in Eq.(\ref{ekul09}) takes the following
form:
\bea
\label{ekul10}
\Pi_{\mu\nu} \es
{\lambda_{\mbox{\tiny${3\over 2}^+$}}^2 \over (m_{\mbox{\tiny${3\over
2}^+$}}^2-p_1^2) (m_{\mbox{\tiny${3\over 2}^+$}}^2-p_2^2)} 
(\rlap/{p}_2 + m_{\mbox{\tiny${3\over
2}^+$}}) \Big[ - g_{\mu\nu} \rlap/{\varepsilon} (f_1^+ +f_2^+)\Big]
(\rlap/{p}_1 + m_{\mbox{\tiny${3\over 2}^+$}}) \nnb \\
\ar {\lambda_{\mbox{\tiny${3\over 2}^+$}} \lambda_{\mbox{\tiny${3\over
2}^-$}}\over (m_{\mbox{\tiny${3\over 2}^-$}}^2-p_1^2)
(m_{\mbox{\tiny${3\over 2}^+$}}^2-p_2^2)}
(\rlap/{p}_2 + m_{\mbox{\tiny${3\over 2}^+$}})
\Big[ - g_{\mu\nu} \rlap/{\varepsilon} (f_1^{tr} +f_2^{tr})\Big]
(\rlap/{p}_1 - m_{\mbox{\tiny${3\over 2}^-$}}) \nnb \\
\ar {\lambda_{\mbox{\tiny${3\over 2}^-$}} \lambda_{\mbox{\tiny${3\over
2}^+$}}\over (m_{\mbox{\tiny${3\over 2}^+$}}^2-p_1^2)
(m_{\mbox{\tiny${3\over 2}^-$}}^2-p_2^2)}
(\rlap/{p}_2 - m_{\mbox{\tiny${3\over 2}^-$}})
 \Big[ - g_{\mu\nu} \rlap/{\varepsilon} (f_1^{{tr}^\ast} +f_2^{{tr}^\ast})\Big]
(\rlap/{p}_1 + m_{\mbox{\tiny${3\over 2}^+$}}) \nnb \\
\ar {\lambda_{\mbox{\tiny${3\over 2}^-$}}^2 \over (m_{\mbox{\tiny${3\over
2}^-$}}^2-p_1^2) (m_{\mbox{\tiny${3\over 2}^-$}}^2-p_2^2)} 
(\rlap/{p}_2 - m_{\mbox{\tiny${3\over
2}^-$}}) \Big[ - g_{\mu\nu} \rlap/{\varepsilon} (f_1^- +f_2^-)\Big]
(\rlap/{p}_1 - m_{\mbox{\tiny${3\over 2}^-$}})~.
\eea

It follows from Eq.(\ref{ekul10}) that only the last term describes the
magnetic moment of the negative parity baryons. Denoting by,
\bea
\label{ekul11}
A \es {\lambda_{\mbox{\tiny${3\over 2}^+$}}^2 (f_1^+ + f_2^+) \over
(m_{\mbox{\tiny${3\over 2}^+$}}^2-p_1^2)
(m_{\mbox{\tiny${3\over 2}^+$}}^2-p_2^2)} \nnb \\
B \es {\lambda_{\mbox{\tiny${3\over 2}^+$}} \lambda_{\mbox{\tiny${3\over
2}^-$}} (f_1^{tr} + f_2^{tr}) \over (m_{\mbox{\tiny${3\over 2}^-$}}^2-p_1^2)
(m_{\mbox{\tiny${3\over 2}^+$}}^2-p_2^2)} \nnb \\
C \es {\lambda_{\mbox{\tiny${3\over 2}^-$}} \lambda_{\mbox{\tiny${3\over
2}^+$}} (f_1^{{tr}^\ast} + f_2^{{tr}^\ast}) \over
(m_{\mbox{\tiny${3\over 2}^+$}}^2-p_1^2)                 
(m_{\mbox{\tiny${3\over 2}^-$}}^2-p_2^2)} \nnb \\
D \es {\lambda_{\mbox{\tiny${3\over 2}^-$}}^2 (f_1^- +f_2^-)
\over (m_{\mbox{\tiny${3\over
2}^-$}}^2-p_1^2) (m_{\mbox{\tiny${3\over 2}^-$}}^2-p_2^2)}~. 
\eea
On the hadronic side, for the invariant function with the structure
$g_{\mu\nu}$, we have
\bea
\label{ekul12}
&&\Big[ A (\rlap/{p}_2 + m_{\mbox{\tiny${3\over 2}^+$}})
(-\rlap/{\varepsilon}) (\rlap/{p}_1 + m_{\mbox{\tiny${3\over 2}^+$}}) +
B (\rlap/{p}_2 + m_{\mbox{\tiny${3\over 2}^+$}}) (\rlap/{\varepsilon})
(\rlap/{p}_1 - m_{\mbox{\tiny${3\over 2}^-$}}) \nnb \\
\ar C(\rlap/{p}_2 - m_{\mbox{\tiny${3\over 2}^-$}})
(\rlap/{\varepsilon})(\rlap/{p}_1 + m_{\mbox{\tiny${3\over 2}^+$}}) +
D (\rlap/{p}_2 - m_{\mbox{\tiny${3\over 2}^-$}}) (\rlap/{\varepsilon}) 
(\rlap/{p}_1 - m_{\mbox{\tiny${3\over 2}^-$}})\Big]~,
\eea
where $p_2=p$ and $p_1=p+q$. Note that in determination of the magnetic
moments we need the values of the form factors only at $q^2=0$. 

Now let us turn our attention to the calculation of the correlator function
from the QCD side. For this goal, as has already been noted, we need the
form of the interpolating currents of the excited state baryons. 
The interpolating currents for the spin-3/2, positive parity baryons are
given as \cite{Rkul15},
\bea
\label{ekul13}
\eta_\mu^{p\ast} \es \varepsilon^{abc} \Big[ \Big( u^{aT} C
\sigma_{\rho\lambda} d^b\Big) \sigma^{\rho\lambda} \gamma_\mu u^c -
\Big( u^{aT} C \sigma_{\rho\lambda} u^b\Big) \sigma^{\rho\lambda}
\gamma_\mu d^c \Big]~, \nnb \\
\eta_\mu^{n\ast} \es \eta_\mu^{p\ast} ( u \lrar d)~, \nnb \\
\eta_\mu^{\Sigma^{\ast+}} \es \varepsilon^{abc} \Big[ \Big( u^{aT} C
\sigma_{\rho\lambda} s^b\Big) \sigma^{\rho\lambda} \gamma_\mu u^c -
\Big( u^{aT} C \sigma_{\rho\lambda} u^b\Big) \sigma^{\rho\lambda}
\gamma_\mu s^c \Big]~, \nnb \\
\eta_\mu^{\Sigma^{\ast-}} \es {\Sigma^{\ast+}} ( u \lrar d)~, \nnb \\
\eta_\mu^{\Sigma^{\ast0}} \es {\varepsilon^{abc} \over \sqrt{2}}
\Big[ \Big( u^{aT} C \sigma_{\rho\lambda} s^b\Big) \sigma^{\rho\lambda}
\gamma_\mu d^c - \Big( u^{aT} C \sigma_{\rho\lambda} d^b\Big)
\sigma^{\rho\lambda} \gamma_\mu s^c \nnb \\
\ar \Big( d^{aT} C \sigma_{\rho\lambda} s^b\Big) \sigma^{\rho\lambda}
\gamma_\mu u^c - \Big( d^{aT} C \sigma_{\rho\lambda} u^b\Big)
\sigma^{\rho\lambda} \gamma_\mu s^c \Big] \nnb \\
\eta_\mu^{\Xi^{\ast 0}} \es \varepsilon^{abc} \Big[ \Big( s^{aT} C
\sigma_{\rho\lambda} u^b\Big) \sigma^{\rho\lambda} \gamma_\mu s^c -
\Big( s^{aT} C \sigma_{\rho\lambda} s^b\Big) \sigma^{\rho\lambda}
\gamma_\mu u^c \Big]~, \nnb \\ 
\eta_\mu^{\Xi^{\ast -}} \es \eta_\mu^{\Xi^{\ast 0}} ( u \lrar d)~.
\eea

As an example we present the result for the correlation function for the
$\Sigma^{\ast+}$ baryon from the QCD side, which can be written as:
\bea
\label{ekul14}
\Pi_{\mu\nu}^{\Sigma^{\ast+}} =
\int d^4x e^{ipx} \lla \gamma (q) \vel \eta_\mu^{\Sigma^{\ast+}} (x)
\bar{\eta}_\nu^{\Sigma^{\ast+}} (0) \ver 0 \rra ~,
\eea
where $a,b,c,a^\prime,b^\prime,c^\prime$ are the color indices, $S_q$ is the
light quark propagator. In further numerical calculations involving excited
baryons containing strange quark, we take into account only linear terms in
strange quark mass $m_s$. The results for the $\Sigma^{\ast-}$, $\Sigma^{\ast0}$,
$\Xi^{\ast 0}$ and $\Xi^{\ast -}$ excited baryons can be obtained from Eq.
(\ref{ekul14}) with the help of the following replacements,
\bea
\label{ekul15}
\Pi_{\mu\nu}^{\Sigma^{\ast-}} \es \Pi_{\mu\nu}^{\Sigma^{\ast+}}
( u \lrar d)~, \nnb \\ 
\Pi_{\mu\nu}^{\Sigma^{\ast0}} \es 
{1\over 2} \Big( \Pi_{\mu\nu}^{\Sigma^{\ast+}} +
\Pi_{\mu\nu}^{\Sigma^{\ast-}} \Big)~, \nnb \\  
\Pi_{\mu\nu}^{\Xi^{\ast0}} \es \Pi_{\mu\nu}^{\Sigma^{\ast+}} 
( u \lrar s)~, \nnb \\  
\Pi_{\mu\nu}^{\Xi^{\ast-}} \es \Pi_{\mu\nu}^{\Xi^{\ast0}}
( u \lrar d)~.  
\eea
In order to calculate the correlation function from the QCD side, we need to
know explicit expression of the light quark propagators, which  
has the following form:
\bea
\label{ekul16}
S_q(x) \es {i \rlap/{x} \over 2 \pi^2 x^4} - {m_q \over 4 \pi^2 x^2} - {\qq
\over 12} \Bigg( 1- {im_q \over 4} \rlap/{x} \Bigg) - {x^2 \over 192} m_0^2
\qq \Bigg( 1- {im_q \over 6} \rlap/{x} \Bigg) \nnb \\
\ek ig_s \int_0^1 dv \Bigg[ {\rlap/{x} \over 16 \pi^2 x^2} G_{\mu\nu}(vx)
\sigma^{\mu\nu} - v x^\mu G_{\mu \nu} (vx) \gamma^\nu {i\over 4 \pi^2 x^2}
\nnb \\
\ek {im_q \over 32 \pi^2} G_{\mu\nu}(vx) \sigma^{\mu\nu} \Bigg(\ln{-x^2\Lambda^2
\over 4} + 2 \gamma_E \Bigg) \Bigg]~,
\eea
where $\Lambda$ is the energy scale in order to separate the perturbative
and nonperturbative sectors. We should note that in numerical calculations
two-gluon and four-quark operators are neglected due to their small
contributions.

In order to obtain perturbative contributions
it is enough to replace one of the free quark operators
(the first two terms in Eq. (\ref{ekul16}) ) by,
\bea
\label{ekul17}
S = -{1\over 2} \int dy y_\nu {\cal F}^{\mu\nu} S^{free} (x-y) \gamma_\mu
S^{free}(y)~,
\eea
where ${\cal A}_\mu = {\ds 1\over 2} {\cal F}^{\mu\nu} y_\nu$ satisfying
$x_\mu {\cal A}^\mu = 0$ in the Schwinger gauge, and ${\cal F}^{\mu\nu}$ is
the electromagnetic field strength tensor, $S^{free}$ is the free quark
propagator. The remaining two
propagators are taken as are given in Eq. (\ref{ekul16}). The nonperturbative
contribution describing the case when a photon interacts with a quark field
nonperturbatively can again be obtained by replacing one of the propagators
by,
\bea
S_{\rho\sigma}^{ab} = -{1\over 4} \Big[\bar{q}^a \Gamma_j q^b
\Gamma_j\Big]_{\rho\sigma}~, \nnb
\eea
where $\Gamma_j$ are the full set of Dirac matrices, and the remaining two
propagators are taken from Eq. (\ref{ekul16}).

These contributions, obviously,  are described by the matrix elements of the
nonlocal operators $\bar{q}\Gamma_j q$ and $\bar{q} G_{\mu\nu} \Gamma_j q$
between the vacuum and photon states. These matrix elements are described in
terms of the photon distribution amplitudes (DAs), whose explicit
expressions are given in \cite{Rkul16}.

Using Eq. (\ref{ekul14}) and the explicit expression for the light quark
operator given in Eq. (\ref{ekul16}), and the definitions of the nonlocal
quark operators between the photon and vacuum states in terms of the photon
DAs, we can perform the calculation for the theoretical part of the
correlation function. It follows from Eq. (\ref{ekul12}) that, in order to
determine the coefficient $D$, which contain the magnetic moment of the
negative parity, spin-3/2 baryons, we need four equations. For this aim we
choose the structures $g_{\mu\nu}(\varepsilon p)$, $(\varepsilon p)
\rlap/{p} g_{\mu\nu}$, $\rlap/{\varepsilon} \rlap/{p} g_{\mu\nu}$ and
$\rlap/{\varepsilon} g_{\mu\nu}$. Denoting the invariant functions of these
structures by $\Pi_1$, $\Pi_2$, $\Pi_3$ and $\Pi_4$, respectively, we get
the following four equations,
\bea
\label{nolabe03}
-2 m_+ (A+B) + 2 m_- (C+D) \es \Pi_1 \nnb \\
-2 (A+B+C+D) \es \Pi_2 \nnb \\
(B-C) (m_+ + m_-) \es \Pi_3 \nnb \\
(C m_- + B m_+) (m_++m_-) \es \Pi_4~.
\eea
Solving these equations for $D$, we get
\bea
\label{ekul18}
D = {1\over 2 (m_+ + m_-)^2} \Big[ (m_+ + m_-) \Pi_1 - m_+ \Pi_2 + 2 m_+
\Pi_3 - 2 \Pi_4 \Big]~.
\eea
The final step for obtaining the sum rules for the magnetic moments is
performing double Borel transformation with respect to the variables
$-(p+q)^2$ and $-p^2$, and subtracting higher states and continuum
contributions. Performing all these operations in the appropriate order, for
the magnetic moment of negative parity baryons we get,
\bea
\label{ekul19}
{\mu \over 3} \es {e^{m_-^2/M^2} \over \lambda_-^2} {1\over 2(m_+ + m_-)^2}
\Big[ (m_+ + m_-)\Pi_1^B - m_+ \Pi_2^B + 2 m_+ \Pi_3^B - 2 \Pi_4^B \Big]~,
\eea
where we take $M_1^2 = 2 M^2$, $M_2^2 = 2 M^2$, with $M^2$ is being the Borel mass
parameter; and $\Pi_i^B$ are the expressions of the corresponding invariant
functions after performing the Borel transformation. The expressions of
$\Pi_i^B$ quite lengthy, and for this reason we do not present their
explicit expressions.

Having obtained the final result for determination of the magnetic moment of
the negative parity, spin-3/2 baryons, which is given in Eq. (\ref{ekul18}),
we now need to calculate their overlap amplitudes (residues). These residues
are determined from from the two-point correlation function given as,
\bea
\label{ekul20}
\Pi_{\mu\nu}(p^2) = i \int d^4x e^{ipx} \lla 0 \vel \mbox{\rm T} \Big\{
\eta_\mu^B (x) \bar{\eta}_\nu^B(0) \Big\}\ver 0 \rra_.
\eea
Saturating this correlation function with positive and negative baryons, and
choosing the structures $g_{\mu\nu} \rlap/{p}$ and $g_{\mu\nu}$ that contain
only spin-3/2 baryons, we get,
\bea
\label{ekul21}
{\lambda_+^2 \over m_+^2-p^2} + {\lambda_-^2 \over m_-^2-p^2} \es T_1 \nnb \\
{\lambda_+^2 m_+\over m_+^2-p^2} + {\lambda_-^2 m_- \over m_-^2-p^2} \es
T_2~,
\eea
where $T_1$ and $T_2$ are the invariant functions on the theoretical side,
multiplying the structures  $g_{\mu\nu} \rlap/{p}$ and $g_{\mu\nu}$,
respectively. Performing Borel transformation over the variable $-p^2$, and
continuum subtraction procedure, for the mass and residues of the negative
parity baryons we get,
\bea
\label{nolabel04}
m_-^2 \es {{\ds d\over \ds d(1/M^2)} \Big[m_+ T_1^B - T_2^B\Big]  
\over m_+ T_1^B - T_2^B}~, \nnb \\
\lambda_-^2 \es { \ds e^{m_-^2/M^2} \over m_+ + m_-} \Big[m_+ T_1^B -
T_2^B\Big]~. \nnb
\eea
The explicit expressions of $T_1$ and $T_2$ for the excited spin-3/2 baryons
are calculated in \cite{Rkul15}.

\section{Numerical calculations}

This section is devoted to the numerical analysis of the sum rules derived
for the negative parity, spin-3/2 baryons. The values of the input
parameters which enter to the sum rules for the magnetic moments are:
$\uu (\mu=1~GeV) = \dd (\mu=1~GeV) = -(0.243)^3~GeV^3$, $\sp(\mu=1~GeV) =
0.8 \uu (\mu=1~GeV)$, $f_{3\gamma} =-0.039$ \cite{Rkul16}, $\Lambda=(0.5\pm
1.0)~GeV$ \cite{Rkul17}, $m_s(\mu=2~GeV) = 111 \pm 6~MeV$ \cite{Rkul18},
and the magnetic susceptibility $\chi(\mu=1~GeV) = -2.85 \pm 0.5~GeV^{-2}$
\cite{Rkul19}.

The main ingredient of the light cone sum rules are the DAs. In our problem
we need the photon DAs, whose analytic expressions are presented in
\cite{Rkul16}. In addition to the above-presented input parameters and
photon DAs, sum rules involve the Borel mass parameter $M^2$ and
continuum threshold $s_0$ as well. Since these two parameters are the
auxiliary ones, the magnetic moments should be independent of them. The
continuum threshold is not totally arbitrary, and it is correlated to the
energy of the first excited states. It is usually chosen in the region $(m_-
+ 0.4)^2 \le s_0 \le (m_- + 0.5)^2~GeV^2$. The working region of the Borel
mass parameter $M^2$ is determined in the following way.
In order to obtain the upper bound of $M^2$ we require that the continuum
and higher states contributions are less than, say, 30\% of the perturbative
contributions. The lower bound of $M^2$ is determined from the requirement
that the contribution of the highest power of $1/M^2$ terms should be less
than 25\% of the highest $M^2$ contributions. These two requirements lead
the following results for the working regions of $M^2$:
\bea
\label{nolabel05}
&&1 \le M^2 \le 3~GeV^2~\mbox{\rm for $p^\ast$, $n^\ast$ and
$\Sigma^\ast$},\nnb \\
&&1.5 \le M^2 \le 3.5~GeV^2~,\mbox{\rm for $\Xi^\ast$}.
\eea
In these regions of the Borel mass parameter $M^2$, the results for the
magnetic moments of the negative parity, spin-3/2 baryons are very weakly
dependent on $M^2$ and $s_0$. As an example, we present in Fig. (1) the
dependence of the magnetic moment for the $p^\ast (3/2)$ state
on $M^2$, at three
fixed values of the continuum threshold $s_0$. We deduce from this figure
that $\mu_{p^\ast} = (1.2 \pm 0.2) \mu_N$.

The results for the other excited states of negative parity, spin-3/2
baryons are presented in Table 1.

% .........................................................

\begin{table}[h]

\renewcommand{\arraystretch}{1.3}
\addtolength{\arraycolsep}{-0.5pt}
\small
$$
\begin{array}{|l|c|}
\hline \hline   
\mu_{p^\ast}         & (1.2 \pm 0.2)\mu_N \\
\mu_{n^\ast}         & (0.9 \pm 0.1)\mu_N \\
\mu_{\Sigma^{+\ast}} & (1.2 \pm 0.2)\mu_N \\
\mu_{\Sigma^{-\ast}} & (-1.5 \pm 0.1)\mu_N \\
\mu_{\Sigma^{0\ast}} & (-0.22 \pm 0.02)\mu_N \\
\mu_{\Xi^{0\ast}}    & (0.36 \pm 0.06)\mu_N \\
\mu_{\Xi^{-\ast}}    & (-1.8 \pm 0.3)\mu_N \\ 
\hline \hline
\end{array}
$$
\caption{}
\renewcommand{\arraystretch}{1}
\addtolength{\arraycolsep}{-1.0pt}
\end{table}

We see from this table that, the value of the magnetic moments of
$\Sigma^{+\ast}$ and $p^{\ast}$, as well as, $\Xi^{-\ast}$ and
$\Sigma^{-\ast}$ are very close to each other, which follows from 
the small difference in magnetic moments due to the SU(3) symmetry
breaking effects. Moreover, in exact SU(3) symmetry, the values of
the magnetic moments of $\Sigma^{0\ast}$ and $\Xi^{0\ast}$ should be equal
to zero. Indeed, our results show that they have quite small values, and
these small nonzero values of the magnetic moments of the relevant baryons  
can be attributed to the SU(3) symmetry breaking. It should finally be
mentioned here that, in exact SU(3) symmetry limit, the relation
\bea
\label{nolabel05}
\mu_{\Sigma^{0\ast}} = {1\over 2} \Big(\mu_{\Sigma^{+\ast}} +
\mu_{\Sigma^{-\ast}} \Big)~ \nnb
\eea
should be satisfied, which is also confirmed by our results. The apparent
small deviation can again be attributed to the SU(3) symmetry breaking
effects.

In conclusion, the magnetic moments of the negative parity, spin-3/2 baryons
are estimated within the LCSR. Checking our predictions in future
experiments could be very useful for understanding the dynamics of the
negative parity baryons.

\newpage

\section*{Figure captions}
{\bf Fig. (1)} The dependence of the magnetic moment for the $p^\ast (3/2)$ state
on $M^2$, at three fixed values of the continuum threshold $s_0$.

\newpage

\begin{figure}
\vskip 3. cm
    \includegraphics{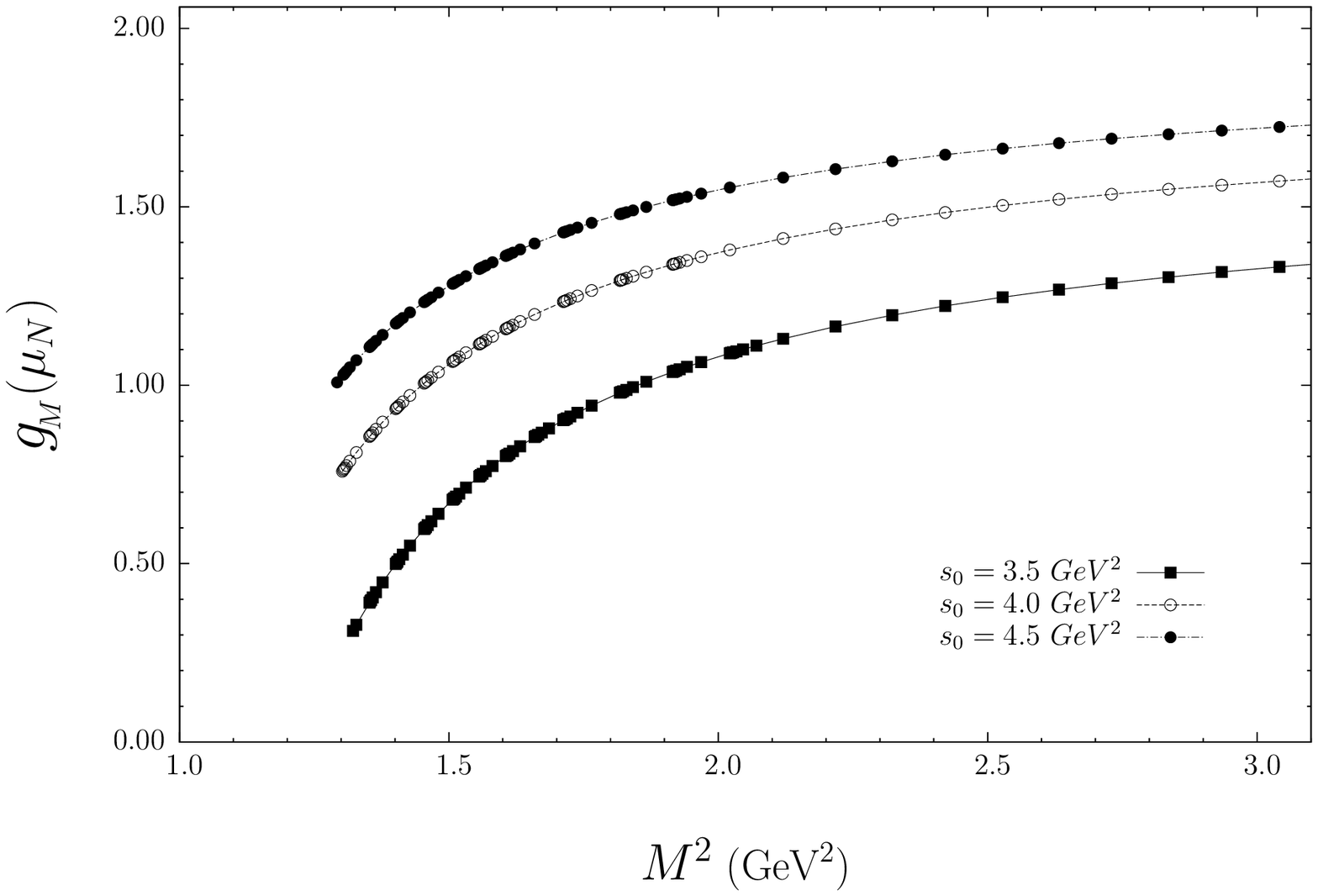}
\vskip 7.0cm
\caption{}
%\begin{center}
%{\bf Fig. 1--a}
%\end{center}
\end{figure}

\end{document}